\documentclass[sigconf,screen]{acmart}
\AtBeginDocument{%
  \providecommand\BibTeX{{%
    \normalfont B\kern-0.5em{\scshape i\kern-0.25em b}\kern-0.8em\TeX}}}

\copyrightyear{2023}
\acmYear{2023}
\setcopyright{acmlicensed}
\acmConference[ESEC/FSE '23]{Proceedings of the 31st ACM Joint European Software Engineering Conference and Symposium on the Foundations of Software Engineering}{December 3--9, 2023}{San Francisco, CA, USA}
\acmBooktitle{Proceedings of the 31st ACM Joint European Software Engineering Conference and Symposium on the Foundations of Software Engineering (ESEC/FSE '23), December 3--9, 2023, San Francisco, CA, USA}
\acmPrice{15.00}
\acmDOI{10.1145/3611643.3616244}
\acmISBN{979-8-4007-0327-0/23/12}
\acmSubmissionID{fse23main-p33-p}
\received{2023-02-02}
\received[accepted]{2023-07-27}

\usepackage{microtype}
\usepackage{booktabs} 
\usepackage{multirow}
\usepackage{tabularx}
\usepackage[T1]{fontenc}
\usepackage{subcaption}
\usepackage{framed}
\usepackage{arydshln}
\usepackage{xcolor}
\usepackage[most]{tcolorbox}
\definecolor{dark-red}{RGB}{255,0,0}
\definecolor{dark-green}{RGB}{0,200,0}

\newcommand{\tmargin}{2pt}
\newcommand{\tsep}{2pt}

\newcommand{\chunk}[2]{%
	\fcolorbox{black}{yellow}{\bfseries\sffamily\scriptsize#1}%
   {$\blacktriangleright$#2$\blacktriangleleft$}%
}

\newcommand{\kisub}[1]{\chunk{Kisub}{{\textcolor{cyan}{\textsl{#1}}}}}

\begin{document}

\title[On the Usage of Continual Learning for Out-of-Distribution Generalization in Pre-trained Language Models ...]{On the Usage of Continual Learning for Out-of-Distribution Generalization in Pre-trained Language Models of Code}



\author{Martin Weyssow}
\affiliation{%
  \institution{DIRO, University of Montreal}
  \city{Montreal}
  \country{Canada}}
\email{martin.weyssow@umontreal.ca}

\author{Xin Zhou}
\affiliation{%
  \institution{Singapore Management University}
  \country{Singapore}}
\email{xinzhou.2020@phdcs.smu.edu.sg}

\author{Kisub Kim}
\affiliation{%
  \institution{Singapore Management University}
  \country{Singapore}}
  \authornote{Corresponding author.}
\email{kisubkim@smu.edu.sg}

\author{David Lo}
\affiliation{%
  \institution{Singapore Management University}
  \country{Singapore}}
\email{davidlo@smu.edu.sg}

\author{Houari Sahraoui}
\affiliation{%
  \institution{DIRO, University of Montreal}
  \city{Montreal}
  \country{Canada}}
\email{sahraouh@iro.umontreal.ca}

\renewcommand{\shortauthors}{Weyssow, et al.}

\begin{abstract}
Pre-trained language models (PLMs) have become a prevalent technique in deep learning for code, utilizing a two-stage pre-training and fine-tuning procedure to acquire general knowledge about code and specialize in a variety of downstream tasks. However, the dynamic nature of software codebases poses a challenge to the effectiveness and robustness of PLMs.
In particular, world-realistic scenarios potentially lead to significant differences between the distribution of the pre-training and test data, \textit{i.e.,} distribution shift, resulting in a degradation of the PLM's performance on downstream tasks. In this paper, we stress the need for adapting PLMs of code to software data whose distribution changes over time, a crucial problem that has been overlooked in previous works. The motivation of this work is to consider the PLM in a non-stationary environment, where fine-tuning data evolves over time according to a software evolution scenario. Specifically, we design a scenario where the model needs to learn from a stream of programs containing new, unseen APIs over time. We study two widely used PLM architectures, \textit{i.e.,} a GPT2 decoder and a RoBERTa encoder, on two downstream tasks, API call and API usage prediction. We demonstrate that the most commonly used fine-tuning technique from prior work is not robust enough to handle the dynamic nature of APIs, leading to the loss of previously acquired knowledge \textit{i.e.,} catastrophic forgetting. To address these issues, we implement five continual learning approaches, including replay-based and regularization-based methods. Our findings demonstrate that utilizing these straightforward methods effectively mitigates catastrophic forgetting in PLMs across both downstream tasks while achieving comparable or superior performance.
\end{abstract}




\maketitle

\section{Introduction}
Prior research~\cite{feng2020codebert,chen2021evaluating,zeng2022extensive} on code representation learning leverages a ubiquitous two-stage procedure to effectively train and specialize pre-trained language models (PLMs) for code-related downstream tasks. The first stage, \textit{i.e.,} the pre-training, involves optimizing the model using self-supervised learning on a large dataset to acquire general knowledge about code. This pre-training phase allows the model to adapt to downstream tasks in the second stage, \textit{i.e.,} the fine-tuning. Previous studies~\cite{feng2020codebert, zhou2021assessing, ahmad2021unified} typically leverage classical transfer learning methods, which consist of "transferring" the pre-trained knowledge to the target task by fine-tuning the model on a task-specific loss function and data.  This approach has been successful in the fields of natural language processing (NLP)~\cite{brown2020language, devlin2018bert} and deep learning for code~\cite{feng2020codebert, chen2021evaluating}.

In this perspective, previous works~\cite{ciniselli2021empirical, watson2022systematic} have primarily focused on stationary settings, neglecting the practical need for models to adapt to changing environments and data over time. Most prior research~\cite{feng2020codebert, guo2022unixcoder, ahmad2021unified} has suggested using transfer learning to fine-tune the model in static environments rather than addressing the dynamic nature of real-world scenarios. In practice, programming languages, software libraries and APIs are prone to change and evolution~\cite{nita2010using, proksch2016evaluating, hoang2020cc2vec}, leading to shifts in the distribution of the underlying software data over time, it is also known as concept drift~\cite{widmer1996learning, lu2018learning}. 
By ignoring the actual evolution of software codebases, existing studies~\cite{watson2022systematic, chen2021evaluating} have focused on fine-tuning and testing pre-trained models of code using stationary datasets.
In practice, the software evolution potentially leads to a noticeable difference between training and test data, \textit{i.e.,} distribution shift, that is often not present in these stationary datasets.
This phenomenon also occurs when the model is put into production and has to deal with real-world data~\cite{hellendoorn2019code, aye2021learning}. 
We argue that creating datasets that reflect real-world software evolution scenarios and distribution shifts is crucial in order to properly evaluate the \textbf{out-of-distribution (OOD) generalization} capability of code models~\cite{shen2021towards}.
The OOD generalization measures a model's ability to generalize to new, unseen data with a significantly different distribution from the training data. Therefore, evaluating how PLMs of code generalize to OOD software data in software evolution scenarios appears as a prime issue.

\begin{figure}[!t]
    \centering
    \includegraphics[width=.9\linewidth]{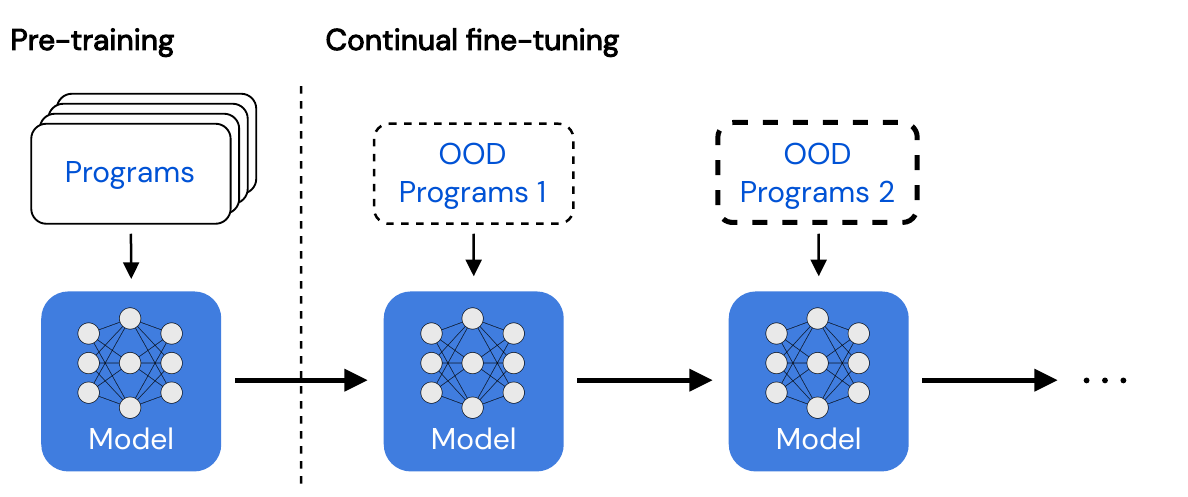}
    \vspace{-1em}
    \caption{Continual fine-tuning of a pre-trained language model of code. After pre-training, the model needs to adapt to new out-of-distribution (OOD) program data over time.}
    \label{fig:intro}
\end{figure}

Existing works on OOD generalization designed the datasets based on various distribution shifts in source code data~\cite{hajipour2022simscood, hu2022codes}. However, they did not address the problem of continually adapting a pre-trained model of code to streams of OOD data. 
The prime goal of our study is to explore methods for a model to better adapt to software evolution scenarios.
In this context, we ask: \textit{how to effectively continually fine-tune a pre-trained model of code to adapt to new data while still considering the past data?} (see Fig.~\ref{fig:intro}). Over the past years, \textbf{continual learning (CL)}~\cite{widmer1996learning, parisi2019continual} has emerged to address this problem, which is relevant to a wide range of research areas, including computer vision~\cite{lomonaco2017core50, baweja2018towards, tao2020few, kirkpatrick2017overcoming} and NLP~\cite{biesialska-etal-2020-continual, thompson2019overcoming, cao2020incremental}. Although transfer learning methods are not tailored for continual learning scenarios, they can still operate to fine-tune a model on streams of data. However, these methods lack robustness, leading to unwanted phenomena such as forgetting past information, known as catastrophic forgetting~\cite{french1999catastrophic, mccloskey1989catastrophic}. There exist other strategies, such as retraining the model from scratch using new data, which are also impractical due to the tremendous computational intensity in the pre-training phase. 
Motivated by these issues of the existing models, we attempt to investigate more robust and scalable fine-tuning techniques.
We hypothesize that continual learning techniques may provide significant benefits over classical transfer learning in this context.

In this paper, we delve into the behavior of PLMs of code in a continual fine-tuning scenario, as depicted in Fig~\ref{fig:intro}. Our objective is twofold: (1) to assess the out-of-distribution generalization capability of PLMs of code and (2) to investigate effective continual fine-tuning strategies to fine-tune the models in the presence of a stream of OOD data.
Specifically, we address these challenges in a scenario reflecting how typical software codebases may evolve in practice. To this end, we create five OOD domain datasets, each introducing new, unseen APIs by the models during their pre-training phase. 
These OOD datasets intend to simulate a stream of data for continual fine-tuning, and each dataset entails a significant distribution shift with respect to the pre-training data. 
As such, our setting establishes an OOD generalization problem. We consider two widely used model architectures: a GPT2-like~\cite{radford2019language} decoder and a RoBERTa-like~\cite{liu2019roberta} encoder pre-trained on code. 
To eliminate any data leakage between the pre-training and fine-tuning data, we decided to pre-train our models from scratch. We do not study the popular existing PLMs like CodeBERT~\cite{feng2020codebert} or CodeT5~\cite{wang2021codet5} because they may be prone to potential data leakage, \textit{i.e.,} seeing the OOD data in pre-training, that we cannot precisely control.
We evaluate the models on two downstream tasks: API call prediction and API usage prediction.
In the first task, the model attempts to predict API calls resulting in a single code token, given code tokens appearing before the call site.
On the other hand, the second task involves the generation of the whole API usage resulting in a sequence of code tokens with the same input format as the prior task. Together, these two tasks provide a comprehensive evaluation of the model's performance in different code generation scenarios.

We start by investigating the impact of OOD data on the performance of the GPT2-like decoder on both downstream tasks in a zero-shot setting, \textit{i.e.,} without fine-tuning the model on the new OOD data.
We find that the model consistently fails to generalize to OOD data by highlighting significant gaps in performance compared to in-distribution data across six evaluation metrics (\textit{e.g.,} up to 75\% drop in BLEU score). 
This finding strongly suggests that pre-training itself is not sufficient and cannot solve OOD generalization in PLMs of code. 
We then evaluate the models' performance in the continual fine-tuning scenario using classical transfer learning and observe notable catastrophic forgetting. 
To address this issue, we implement a straightforward yet computationally inefficient cumulative fine-tuning approach by utilizing a replay buffer of infinite size.
The results show that the approach drastically mitigates forgetting. 
Finally, we compare the performance of classical transfer learning to that of replay-based and regularization-based continual learning methods. 
Replay methods are considered tough-to-beat strategies for continual learning and consist of maintaining a small replay buffer containing samples from previously seen data. During fine-tuning, we use the replay buffer in conjunction with the current OOD training set to fine-tune the PLM. 
We explore regularization-based methods, including EWC~\cite{kirkpatrick2017overcoming}, SI~\cite{zenke2017continual} and RWalk~\cite{chaudhry2018riemannian}, which add regularization terms to the loss function at fine-tuning to prevent extensive changes in important parameters of the PLM. We chose those methods as they are computationally efficient, well-known, and considered strong baselines in the continual learning literature.
We discover that those continual learning methods significantly reduce forgetting while achieving similar or superior effectiveness on both tasks.

To the best of our knowledge, this work constitutes the first initiative to study continual fine-tuning for OOD generalization of PLMs of code.
We believe that the impact of continual learning in this research area has the potential to be far-reaching, particularly due to the inherent evolution of software data over time, and we discuss this aspect in more detail in the discussion section of the paper (see Section~\ref{sec:discussion}). 
Our contributions can be summarized as follows:
\begin{enumerate}
    \item We demonstrate that PLMs of code fail to generalize to OOD data and highlight the need for further investigation in this area.
    
    \item We conduct a study on the behavior of two pre-trained model architectures of code in a continuous learning environment, showing that classical transfer learning lacks robustness and is prone to catastrophic forgetting.
    
    \item We compare five continual learning methods, including replay-based and regularization-based approaches, in our continual fine-tuning scenario. We show the superiority of continual learning over classical transfer learning.
    
    \item We provide a large-scale dataset of Java code snippets and their API usage sequences, including pre-training data and a procedure for extracting OOD data.
\end{enumerate}

\noindent\textbf{Organization.} In Section~\ref{sec:preliminaries}, we discuss preliminaries on continual learning. In Section~\ref{sec:experiments}, we go through our experimental design. We present the results of our experiments in Section~\ref{sec:results}. In Section~\ref{sec:discussion}, we discuss the threats to the validity of our study, as well as potential broader impact and future research directions. We introduce the related work on out-of-distribution generalization and continual learning for pre-trained language models in Section~\ref{sec:related_work}.  Finally, we discuss some future work and conclude this work in Section~\ref{sec:ccl}.

\section{Preliminaries on Continual learning}
\label{sec:preliminaries}
Existing PLMs such as BERT~\cite{devlin2018bert} or GPT~\cite{brown2020language} typically operate in transfer learning settings. By using a two-stage pre-training/fine-tuning procedure, these models can be specialized for a wide range of downstream tasks. However, in this setting, the data used for pre-training or fine-tuning are often assumed to be stationary, which is not reflective of real-world situations. In practice, transfer learning methods can still be applied to non-stationary data, such as a stream of data, but this technique is prone to catastrophic forgetting~\cite{french1999catastrophic,mccloskey1989catastrophic}.

To address the above issues, prior works~\cite{french1999catastrophic, kirkpatrick2017overcoming, li2017learning, hadsell2020embracing, aljundi2018memory, wan2013regularization} introduced the concept of {\em continual learning} and designed specific techniques to mitigate catastrophic forgetting. The primary assumption for continual learning is that the neural network should possess the ability to adapt to new data or tasks while maintaining stability on previous data or tasks, often referred to as the plasticity--stability dilemma. Considering continual learning is particularly interesting for OOD generalization problems, as continual learning methods focus on a keeping good plasticity--stability trade-off. Altogether, it has to potential to enhance the generalizability of PLMs to a broader range of data.
Continual learning methods often operate in constrained scenarios, and Hadsell et al.~\cite{hadsell2020embracing} outline a comprehensive list of objectives to balance in continual learning scenarios. There exist three main categories of methods for continual learning as defined in a previous study~\cite{de2019continual}. \textit{Replay-based methods} store samples from previous experiences, \textit{i.e.,} previous stream of data, in a replay buffer or use generative approaches to generate examples similar to those of previous experiences. The replay buffer is used in conjunction with the current experience data to train the model. Replay-based methods help the network gain stability by enabling the network to train on previous samples, \textit{i.e.,} stored in the replay buffer while adapting to new data.
\textit{Regularization-based methods} add a regularization term to the loss function to prevent catastrophic forgetting by penalizing changes to important neural network parameters. 
Examples of regularization-based methods include EWC~\cite{kirkpatrick2017overcoming}, SI~\cite{zenke2017continual} and RWalk~\cite{chaudhry2018riemannian}.
Finally, \textit{parameter isolation methods} use dynamic architectures to incorporate knowledge from previous experiences to mitigate interference~\cite{rusu2016progressive}.

%


\section{Experimental Design}
\label{sec:experiments}
In this section, we describe the experimental setup of our study. We carefully control our data and model setup to implement our out-of-distribution scenario. 
We first outline the construction of our dataset and the generation of OOD data for continual fine-tuning. Next, we discuss the pre-training procedure of our models, the target downstream tasks and evaluation metrics. We present the results of our experiments in Section~\ref{sec:results}.

\subsection{Dataset Construction}
\label{sec:dataset_construction}
\begin{figure}[!t]
    \centering
    \includegraphics[width=\linewidth]{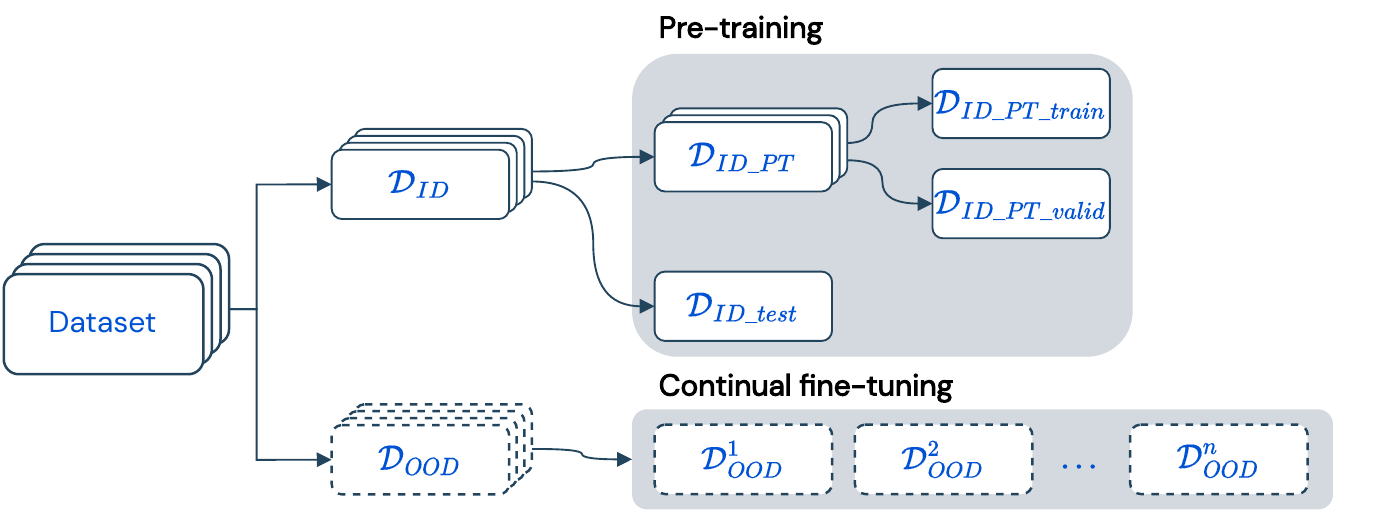}
    \vspace{-0.5em}
    \caption{Procedure to extract the ID data used for model pre-training, and the OOD data used for continual fine-tuning.}
    \label{fig:data_splitting}
\end{figure}
\renewcommand{\arraystretch}{1.2}
\setlength{\arrayrulewidth}{.5pt}
\begin{table*}[!ht]
\centering
\small
\caption{Out-of-distribution dataset details.} 
\vspace{-1em}
\resizebox{1.6\columnwidth}{!}{%
  \begin{tabular}{lcrp{0.40\linewidth}cc}
    \toprule
    \textsc{Dataset} & \multicolumn{1}{r}{Domain} & \multicolumn{1}{c}{Package} & \multicolumn{1}{c}{Interfaces} & \multicolumn{1}{c}{\# train} & \multicolumn{1}{c}{\# test} \\ 
    \midrule
    \multirow{4}{*}{$\mathcal{D}_{OOD}^1$} & \multirow{4}{*}{General} & {\small java.util.concurrent} & {\footnotesize BlockingQueue, ThreadPoolExecutor} & \multirow{4}{*}{47,213} & \multirow{4}{*}{5,239} \\
    & & {\small java.math} & {\footnotesize BigInteger} \\
    & & {\small java.util} & {\footnotesize Base64, TreeSet} \\
    & & {\small java.net} & {\footnotesize ForkJoinPool, Proxy, ServerSocket, SocketAddress, URLEncoder} \\
    \midrule
    $\mathcal{D}_{OOD}^2$ & Security & {\small java.security} & {\footnotesize Cipher, CodeSource, Identity, KeyFactory, KeyPair, MessageDigest, Policy, Provider, Security, Timestamp} & 27,189 & 3,017 \\ 
    \midrule
    \multirow{5}{*}{$\mathcal{D}_{OOD}^3$} & \multirow{5}{*}{Android} & {\small android.view} & {\footnotesize Display, InputEvent, Window} & \multirow{5}{*}{28,400} & \multirow{5}{*}{3,150} \\
    & & {\small android.widget} & {\footnotesize Checkbox, GridLayout} & \\
    & & {\small android.media} & {\footnotesize AudioFormat, ImageReader} & \\
    & & {\small android.hardware} & {\footnotesize Camera, Sensor} & \\
    & & {\small android.database} & {\footnotesize DatabaseUtils} & \\
    \midrule
    $\mathcal{D}_{OOD}^4$ & Web & {\small org.springframework} & {\footnotesize CacheManager, ClassPathResource, DataBuffer, HttpMessage, HttpRequest, JdbcTemplate, MessageChannel, MessageHandler, TaskExecutor} & 16,295 & 1,805 \\ 
    \midrule
    \multirow{5}{*}{$\mathcal{D}_{OOD}^5$} & \multirow{5}{*}{Guava} & {\small com.google.common.graph} & {\footnotesize GraphBuilder, Network} & \multirow{5}{*}{13,448} & \multirow{5}{*}{1,489} \\ 
    & & {\small com.google.common.io} & {\footnotesize ByteSource, ByteStreams} & \\
    & & {\small com.google.common.cache} & {\footnotesize CacheBuilder, LoadingCache} & \\
    & & {\small com.google.common.collect} & {\footnotesize ListMultimap, Multimap} & \\
    & & {\small com.google.common.base} & {\footnotesize CharMatcher, Splitter} & \\
    \bottomrule
  \end{tabular}
  }
\label{tab:ood-dataset}
\end{table*}

Pre-training language models from scratch require a large amount of data for the loss of the model to converge. With that in mind, we constructed our large dataset using programs crawled from GitHub using Google BigQuery\footnote{\url{https://cloud.google.com/bigquery}}. Specifically, we focused on \texttt{Java} programs and began by collecting all \texttt{java} files stored in GitHub repositories. Next, we used Groum~\cite{nguyen2009graph} to extract all methods defined in the \texttt{java} files along with their API usage sequences. We extracted the API usage sequences to facilitate our data splitting and obtain the position of each API site inside the methods to implement our downstream tasks. Each sample consists of all the tokens of a method.
To avoid duplication bias in our experiments~\cite{allamanis2019adverse}, we deduplicated the dataset by comparing the hash of each method. The resulting dataset contains more than 68M \texttt{Java} methods. For our experiments, we shuffled these 68M methods and randomly selected 10M methods to constitute our initial dataset. Fig.~\ref{fig:data_splitting} illustrates how we further split the data for our experiments. Because we chose the pre-train PLMs from scratch, we have to split our data into in-distribution (ID) data, used for model pre-training, and OOD data, used for continual fine-tuning. We also need to properly extract the OOD data to align with our scenario consisting of introducing new, unseen APIs over time to the PLM during fine-tuning. 

\subsubsection*{\textbf{Out-Of-Distribution Dataset -- $\mathcal{D}_{OOD}$}.} 
We create five OOD datasets, $\mathcal{D}_{OOD}^1, ..., \mathcal{D}_{OOD}^5$. Each OOD dataset represents a unique domain that encompasses a high-level functionality of APIs. For example, we have a domain {\em Security} that comprises APIs related to programming security-related code and a domain {\em Guava} that includes only APIs from the Guava\footnote{https://github.com/google/guava} library. To create each OOD dataset, we randomly select 10 interfaces from packages/libraries related to their domain. Finally, we associate to each domain dataset all APIs within the selected interfaces, excluding class construction methods. Table~\ref{tab:ood-dataset} summarizes the dataset $\mathcal{D}_{OOD}$, which contains 147,245 samples in total.

To form each OOD dataset, we select samples from the pool of 10 million Java methods that manipulate at least one of their associated API. 
In our experiments, we perform continual fine-tuning on the training sets associated with the OOD dataset $\mathcal{D}_{OOD}^1, ..., \mathcal{D}_{OOD}^5$ sequentially.
Therefore, to prevent data leakage, we exclude samples that manipulate APIs from multiple domains. This elimination of samples removes a significant threat to the validity of our OOD scenario and ensures that APIs are introduced as intended during the fine-tuning process. To obtain representative test sets, we randomly select 10\% of samples that manipulate each API within each OOD dataset and used the selected samples to form the corresponding domain test set.

\subsubsection*{\textbf{In-Distribution Dataset -- $\mathcal{D}_{ID}$}.} We obtain $\mathcal{D}_{ID}$ by removing the samples in $\mathcal{D}_{OOD}$ from the initial data. Then, we shuffle $\mathcal{D}_{ID}$ and randomly select 50,000 samples for test ($\mathcal{D}_{ID\_test}$). $\mathcal{D}_{ID\_PT}$ contains the remaining samples for pre-training, and we randomly select 100,000 for model validation ($\mathcal{D}_{ID\_PT\_valid}$). In particular, those samples allow us to monitor the evolution of the loss of the model on an independent validation set to avoid overfitting the pre-training data.
In total, the pre-training set $\mathcal{D}_{ID\_PT\_train}$ contains more than 9M samples to pre-train the models.

\subsection{Models and Tasks Setup}
\label{sec:models_and_tasks}

In this work, we consider two widely-used deep learning architectures for code: a RoBERTa-like encoder~\cite{liu2019roberta} and a GPT2-like decoder~\cite{radford2019language}. 
We deliberately exclude the utilization of large language models (LLMs) in our research due to the substantial computational resources essential for their pre-training.
To comprehensively addressour OOD scenario, it is imperative to pre-train a model from scratch prior prior to continually fine-tune it on code containing new, unseen APIs. 
Consequently, we opt to evaluate two smaller models architectures, namely RoBERTa and GPT-2, which either serve as foundational models for PLMs like CodeBERT~\cite{feng2020codebert} or to generative models.


\subsubsection*{\textbf{Decoder -- $\mathcal{M}_{dec}$}.} The decoder model is based on the GPT-2 architecture, with the same hyperparameters, and is pre-trained using a causal language modeling objective, \textit{i.e.,} left-to-right next token prediction. As we conducted our experiments under limited resources, we implemented a small version of GPT-2 with 110 million trainable parameters and pre-train the model for 100,000 steps. We use early stopping to select the best model checkpoint, based on the loss on the validation set $\mathcal{D}_{ID\_PT\_valid}$.

\subsubsection*{\textbf{Encoder -- $\mathcal{M}_{enc}$}.} The encoder model is based on the RoBERTa architecture, with the same hyperparameters, and is pre-trained using a masked language modeling objective. We implemented a base version of RoBERTa. The model has 125 million trainable parameters and is pre-trained similarly to the decoder model, with early stopping used to select the best checkpoint.
Note that conversely to $\mathcal{M}_{dec}$, the encoder's architecture is not suitable for generation tasks. Therefore, we add a randomly initialized language modeling head on top of it for fine-tuning using the OOD datasets. As a result, we expect $\mathcal{M}_{enc}$ to be less stable than $\mathcal{M}_{dec}$ and more prone to catastrophic forgetting since the language modeling head is not pre-trained. This comparison provides valuable insights into the robustness of two different architectures.

\begin{figure}[!t]
    \centering
    \includegraphics[width=\linewidth]{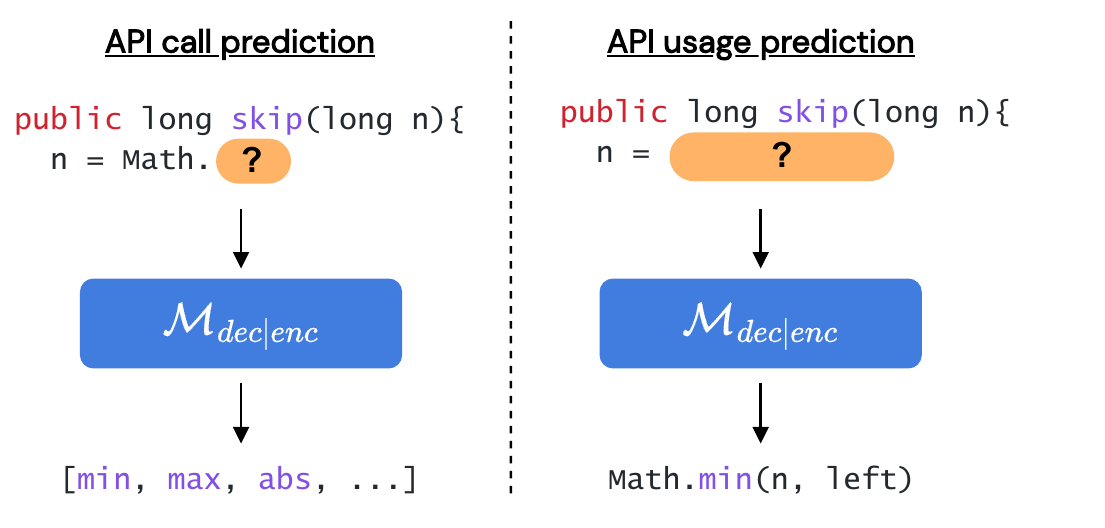}
    \vspace{-2.5em}
    \caption{Overview of the downstream tasks. In the API call prediction task, the model outputs a list of top-$k$ candidates to predict the API call token (\textit{i.e.,} min). In the API usage prediction task, the model attempts to predict all the tokens constituting the API usage (\textit{interface name, method name, parameters and syntactical tokens}). The models only leverage left-context tokens to generate a prediction.}
    \label{fig:tasks}
\end{figure}

\subsubsection*{\textbf{Downstream Tasks}.} We employ two downstream tasks to evaluate the ability of our PLMs of code to learn and adapt to new software data that introduce new, unseen APIs over time. Fig.~\ref{fig:tasks} illustrates both tasks. For API call prediction, the model takes as input all the tokens of the method preceding the call site of the API and generates top-$k$ candidates. For API usage prediction, the model takes as input the same tokens as for the API call prediction task, but attempts to generate the whole API usage (interface name, method name, parameters and syntactical tokens), which constitutes a more challenging task.
The rationale for evaluating the PLMs on these two downstream tasks is to select tasks where prior knowledge about the APIs seems decisive to effectively perform the task. 
Consequently, the choice for these two tasks is highly relevant to our continual OOD scenario, and it allows us to directly measure the impact of OOD APIs on the effectiveness of the PLMs. 
In Section~\ref{sec:broader_impact}, we discuss the applicability of our methodology to other code-related tasks. 

\subsubsection*{\textbf{Evaluation Metrics}.} We measure the performance of the models on both downstream tasks with metrics used in prior works. For API call prediction, we report the Exact Match@k (EM@k), which gives the percentage of correct predictions when considering lists of $k$ candidates. For API usage prediction, we report BLEU score, Exact Match (EM), and CodeBLEU~\cite{ren2020codebleu}. 

To measure how the models perform in a continual learning environment, we use two meta-metrics adapted from prior works~\cite{chaudhry2018riemannian, jie2022alleviating}: the \textit{Average (A)} and \textit{Forgetting (F)} metrics. We define the average $A_M$ of a metric $M$ on a test dataset $\mathcal{D}_{OOD}^i$ as:
$$A_M = \frac{1}{T} \sum_{j=i}^T M_j(\mathcal{D}_{OOD}^i)\:,$$
where $j$ refers to the next incremental learning steps after the $i$-th included. $M_j$ denotes an evaluation metric, \textit{e.g.,} EM@k, computed at time step $j$ on the test set and $T$ denotes the maximum number of fine-tuning steps, \textit{i.e.,} five in our case. The Average metric only gives information on how accurate the model is but does not provide any insight into its ability to mitigate catastrophic forgetting. We define the forgetting $F^k_M$ of a metric $M$ on a test dataset $\mathcal{D}_{OOD}^i$ at time step $k$ as:
$$F^k_M = M_i(\mathcal{D}_{OOD}^i)\:-\:M_k(\mathcal{D}_{OOD}^i)\:,\: i < k\:.$$
This is the difference between the first time the metric is computed, \textit{i.e.,} after fine-tuning the model on $\mathcal{D}_{OOD}^i$ at time step $i$, and the metric computed at time step $k$. $F^k_M$ gives information on the stability of the model, \textit{i.e.,} its capability to not forget from the past. Therefore, the lower $F^k_M$, the better.

\subsubsection*{\textbf{Implementation Details}.} To pre-train $\mathcal{M}_{dec}$ and $\mathcal{M}_{enc}$, we used four Tesla V100-SXM2-32GB GPUs. It took about 7 days to pre-train $\mathcal{M}_{dec}$, and 2 days to pre-train $\mathcal{M}_{enc}$. For fine-tuning and inference, we used a single Tesla V100-SXM2-32GB GPU. We used Huggingface's libraries~\cite{wolf2019huggingface} to implement the models and store the datasets. To implement the continual learning approaches, we used Avalanche~\cite{lomonaco2021avalanche}. We provide all the implementation details of our experiments and release our data publicly in our replication package (see Data Availability section).

\section{Experimental Results}
\label{sec:results}
\subsection{How Does $\mathcal{M}_{dec}$ Generalize to ID and OOD Data in Zero-Shot?}
\label{sec:zero-shot}
\renewcommand{\arraystretch}{1.2}
\setlength{\arrayrulewidth}{.5pt}
\begin{table}[!t]
\centering
\caption{API call prediction results in zero-shot using $\mathcal{M}_{dec}$.}
\vspace{-1em}
    \begin{tabular*}{\linewidth}{l@{\extracolsep{\fill}}*{3}{c}}
    \toprule
    & \multicolumn{3}{c}{\textsc{Metrics}} \\ 
    \cline{2-4} 
    \textsc{Dataset} & \multicolumn{1}{c}{EM@1} & \multicolumn{1}{c}{EM@5} & \multicolumn{1}{c}{EM@10} \\ 
    \midrule
    $\mathcal{D}_{ID\_test}$ & \underline{72.88} & \underline{83.30} & \underline{85.60} \\
    \midrule
    $\mathcal{D}_{OOD}$ & 40.82 (44\% \textcolor{dark-red}{$\downarrow$}) & 51.19 (38.5\% \textcolor{dark-red}{$\downarrow$}) & 54.17 (36.7\% \textcolor{dark-red}{$\downarrow$}) \\ 
    \hdashline
    \hspace{1em} $\mathcal{D}_{OOD}^1$ & 49.91 (31.6\% \textcolor{dark-red}{$\downarrow$}) & 62.0 (25.6\% \textcolor{dark-red}{$\downarrow$}) & 64.46 (24.6\% \textcolor{dark-red}{$\downarrow$}) \\ 
    \hspace{1em} $\mathcal{D}_{OOD}^2$ & 53.72 (26.3\% \textcolor{dark-red}{$\downarrow$}) & 62.59 (24.8\% \textcolor{dark-red}{$\downarrow$}) & 64.93 (24.2\% \textcolor{dark-red}{$\downarrow$}) \\ 
    \hspace{1em} $\mathcal{D}_{OOD}^3$ & 23.78 (67.4\% \textcolor{dark-red}{$\downarrow$}) & 32.64 (60.8\% \textcolor{dark-red}{$\downarrow$}) & 36.33 (57.6\% \textcolor{dark-red}{$\downarrow$}) \\ 
    \hspace{1em} $\mathcal{D}_{OOD}^4$ & 30.72 (57.9\% \textcolor{dark-red}{$\downarrow$}) & 43.67 (47.3\% \textcolor{dark-red}{$\downarrow$}) & 47.89 (44\% \textcolor{dark-red}{$\downarrow$}) \\ 
    \hspace{1em} $\mathcal{D}_{OOD}^5$ & 37.54 (48.6\% \textcolor{dark-red}{$\downarrow$}) & 49.53 (40.6\% \textcolor{dark-red}{$\downarrow$}) & 53.22 (47.9\% \textcolor{dark-red}{$\downarrow$}) \\ 
    \bottomrule
    \end{tabular*}
    \label{tab:zero-shot_api_call}
\end{table}

In this experiment, we evaluate the performance of the model $\mathcal{M}_{dec}$ on the ID and OOD test data in a zero-shot setting for both downstream tasks. We do not experiment with $\mathcal{M}_{enc}$ as the model is not capable of generating code before fine-tuning and, therefore, cannot operate in a zero-shot setting. The purpose of this experiment is twofold. First, it aims to validate the experimental setup of our study. If we observe significant differences in the evaluation metrics obtained on the ID and OOD datasets, it would suggest that our OOD scenario is well-formed and reasonable. Secondly, significant gaps between the ID and OOD test data imply that PLMs such as $\mathcal{M}_{dec}$ still require the use of robust transfer learning or continual learning techniques to generalize to new data without forgetting about past data.

\subsubsection*{\textbf{API Call Prediction}} Table~\ref{tab:zero-shot_api_call} reports the EM@1, EM@5 and EM@10 on the ID and OOD test datasets. The results show that the model performs well on ID data, reaching almost $73\%$ in EM@1. However, when tested on OOD data, the performance drops significantly. The decline in performance is less severe when considering more API call candidates, but it remains a significant issue. Furthermore, variations in the performance decline are observed across different OOD datasets. For example, the model performs better on the Security domain ($\mathcal{D}_{OOD}^2$) than domains such as Android ($\mathcal{D}_{OOD}^3$) or Web ($\mathcal{D}_{OOD}^4$), which likely contain more domain-specific API calls.

\renewcommand{\arraystretch}{1.2}
\setlength{\arrayrulewidth}{.5pt}
\begin{table}[!t]
\centering
\caption{API usage prediction results in zero-shot using $\mathcal{M}_{dec}$.}
\vspace{-1em}
    \begin{tabular*}{\linewidth}{l@{\extracolsep{\fill}}*{3}{c}}
    \toprule
    & \multicolumn{3}{c}{\textsc{Metrics}} \\ 
    \cline{2-4} 
    \textsc{Dataset} & BLEU & EM & CodeBLEU \\ 
    \midrule
    $\mathcal{D}_{ID\_test}$ & \underline{21.19} & \underline{51.54} & \underline{29.94} \\
    \midrule
    $\mathcal{D}_{OOD}$ & 8.57 (59.56\% \textcolor{dark-red}{$\downarrow$}) & 33.74 (34.54\% \textcolor{dark-red}{$\downarrow$}) & 20.03 (33.10\% \textcolor{dark-red}{$\downarrow$}) \\ 
    \hdashline
    \hspace{1em} $\mathcal{D}_{OOD}^1$ & 5.94 (71.97\% \textcolor{dark-red}{$\downarrow$}) & 34.29 (33.47\% \textcolor{dark-red}{$\downarrow$}) & 15.71 (47.53\% \textcolor{dark-red}{$\downarrow$}) \\ 
    \hspace{1em} $\mathcal{D}_{OOD}^2$ & 11.81 (44.27\% \textcolor{dark-red}{$\downarrow$}) & 40.46 (21.50\% \textcolor{dark-red}{$\downarrow$}) & 25.64 (14.36\% \textcolor{dark-red}{$\downarrow$}) \\ 
    \hspace{1em} $\mathcal{D}_{OOD}^3$ & 7.26 (65.74\% \textcolor{dark-red}{$\downarrow$}) & 28.01 (45.65\% \textcolor{dark-red}{$\downarrow$}) & 16.49 (44,92\% \textcolor{dark-red}{$\downarrow$}) \\  
    \hspace{1em} $\mathcal{D}_{OOD}^4$ & 15.55 (26.62\% \textcolor{dark-red}{$\downarrow$}) & 29.39 (42.98\% \textcolor{dark-red}{$\downarrow$}) & 19.72 (34.13\% \textcolor{dark-red}{$\downarrow$}) \\ 
    \hspace{1em} $\mathcal{D}_{OOD}^5$ & 5.11 (75.88\% \textcolor{dark-red}{$\downarrow$}) & 30.71 (40.42\% \textcolor{dark-red}{$\downarrow$}) & 25.81 (13.79\% \textcolor{dark-red}{$\downarrow$}) \\  
    \bottomrule
    \end{tabular*}
    \label{tab:zero-shot_api_usage}
\end{table}

\subsubsection*{\textbf{API Usage Prediction}} Table~\ref{tab:zero-shot_api_usage} reports the BLEU score, EM and CodeBLEU score on both ID and OOD test datasets. The results indicate that the model performs poorly on OOD data in comparison to ID data, with significant decreases in all evaluation metrics. Additionally, we notice that the EM and CodeBLEU metrics vary similarly to the EM@k metrics on the API call prediction task. The Android and Web domains experience the most severe drops, whereas the Security domain experiences the least severe drop. 

\begin{tcolorbox}[tile,size=fbox,boxsep=2mm,boxrule=0pt,top=0pt,bottom=0pt,
borderline west={1mm}{0pt}{blue!50!white},colback=blue!5!white]
 Our results demonstrate that the model $\mathcal{M}_{dec}$ (without fine-tuning) is unable to generalize to OOD data while showing strong performance on ID data. Our findings also support the validity of our OOD dataset as a realistic and meaningful test of the model's ability to adapt to new data in a continuous environment.
\end{tcolorbox}

\subsection{Do Models Forget About Past Data Using Classical Transfer Learning?}
\label{sec:ft-naive}

\begin{figure*}[!t]
    \centering
     \begin{subfigure}[b]{0.32\linewidth}
        \includegraphics[width=\linewidth]{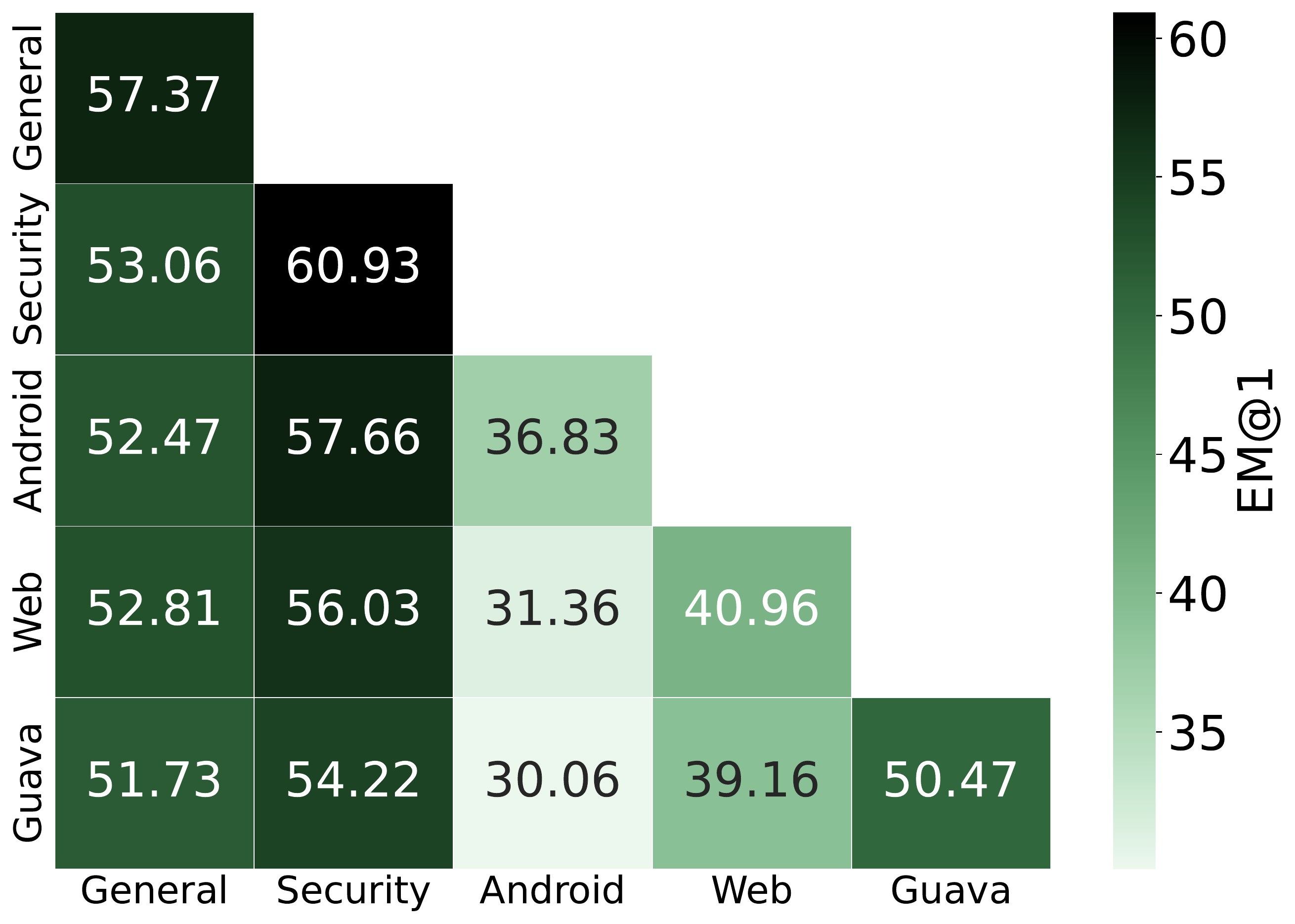}
        \caption{$\mathcal{M}_{dec}$ -- API call prediction.}
        \label{fig:dec_api-call_map_pass@1}
     \end{subfigure}
     \begin{subfigure}[b]{0.32\linewidth}
        \includegraphics[width=\linewidth]{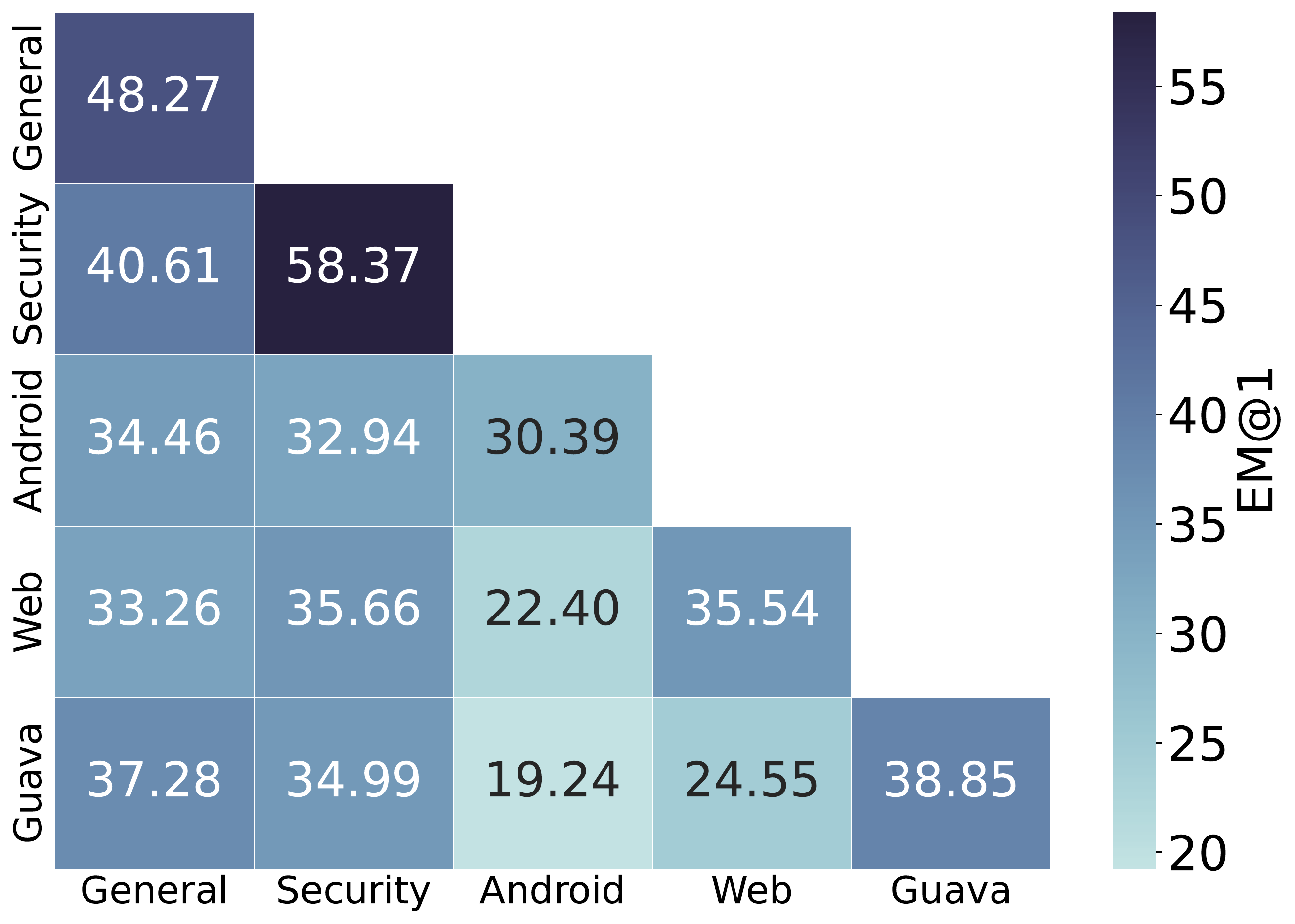}
        \caption{$\mathcal{M}_{enc}$ -- API call prediction.}
        \label{fig:enc_api-call_map_pass@1}
     \end{subfigure}
     \vfill
     \begin{subfigure}[b]{0.32\linewidth}
        \includegraphics[width=\linewidth]{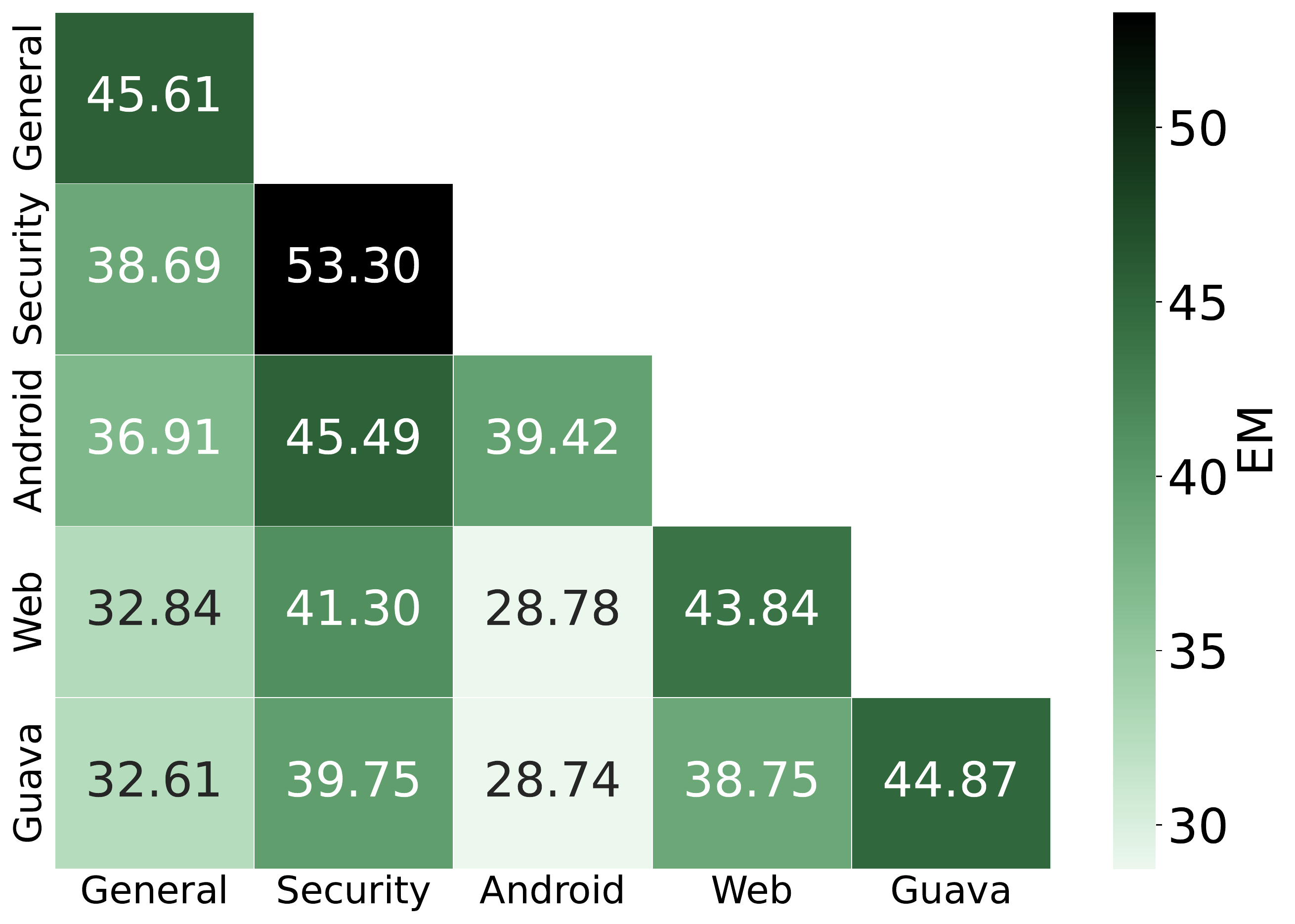}
        \caption{$\mathcal{M}_{dec}$ -- API usage prediction.}
        \label{fig:dec_api-usage_map_em}
     \end{subfigure}
     \begin{subfigure}[b]{0.32\linewidth}
        \includegraphics[width=\linewidth]{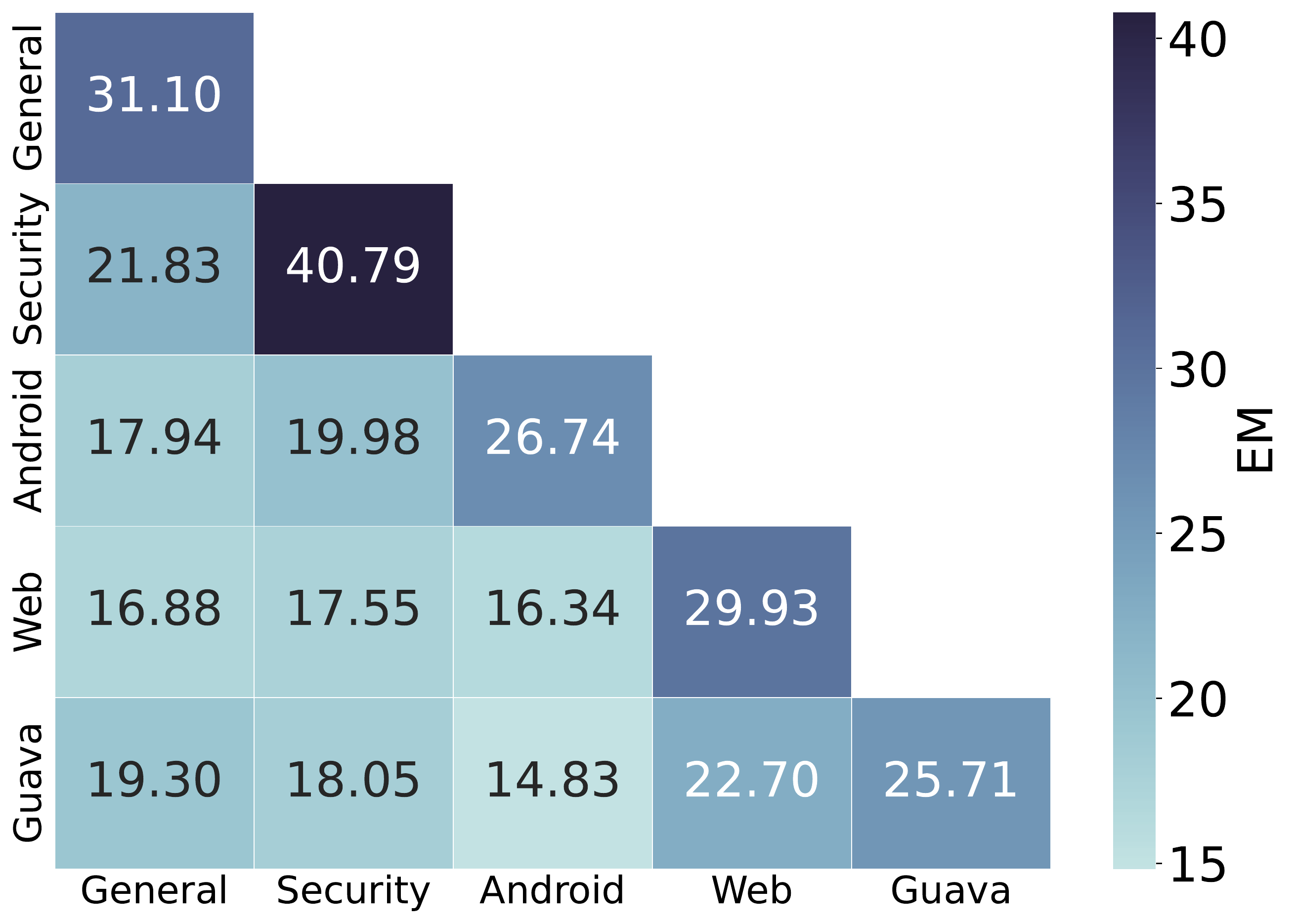}
        \caption{$\mathcal{M}_{enc}$ -- API usage prediction.}
        \label{fig:enc_api-usage_map_em}
     \end{subfigure}
     \vspace{-1em}
     \caption{Naive fine-tuning approach results.}
     \label{fig:naive_finetuning}
\end{figure*}

In this section, we evaluate how classical transfer learning, \textit{i.e.,} using fine-tuning as in prior work, performs in the continual learning scenario. We fine-tune the models $\mathcal{M}_{dec}$ and $\mathcal{M}_{enc}$ sequentially on the stream of OOD datasets $\mathcal{D}_{OOD}^1, ..., \mathcal{D}_{OOD}^5$. We refer to this approach as "naive fine-tuning", a common term used in the continual learning literature to refer to classical transfer learning, as it does not utilize mechanisms to address catastrophic forgetting. We report the results in terms of EM@1 for API call prediction and EM for API usage prediction. Fig.~\ref{fig:naive_finetuning} illustrates the evolution of the EM@1 and EM metrics on the OOD test sets throughout the fine-tuning steps for both models. Each column of a heatmap refers to the evolution of the performance of the model on a particular test set, and each row refers to a new incremental fine-tuning step. Note that we do not compute the metric on a test set whose corresponding training set has not been seen yet by the model. To quantify catastrophic forgetting, we report the Forgetting ($F$) metrics of the EM@1 and EM metrics in Table~\ref{tab:naive-baseline-forgetting}. We do not report all the values for every previously introduced metric as we have a strict page limit, and report them in our replication package.

\subsubsection*{\textbf{Fine-Tuning Details}} At each time step $t$, we fine-tune the models' checkpoints from the previous time step on the dataset $\mathcal{D}_{OOD}^t$. We select 10\% of the training samples from each OOD dataset as a validation set. For each fine-tuning, we set the number of training epochs to 10 and use early stopping by monitoring the evolution of the validation loss with a patience of two epochs. We keep the best checkpoints of the models at each fine-tuning step $t$ and compute the task metrics on the previous and current test sets.

\renewcommand{\arraystretch}{1.1}
\setlength{\arrayrulewidth}{.5pt}
\begin{table}[!t]
\centering
\small
\caption{Forgetting metrics for the naive fine-tuning baseline.} \label{tab:naive-baseline-forgetting}
\vspace{-1em}
    \begin{tabular*}{.7\linewidth}{lr@{\extracolsep{\fill}}*{2}{c}}
    \toprule
    \textsc{Model} & \textsc{Dataset} & $F^5_{\text{EM@1}}$ & $F^5_{\text{EM}}$ \\
    \midrule
    \multirow{4}{*}{$\mathcal{M}_{dec}$} & General {\footnotesize($\Delta t=4$)} & 5.64 & 13.00 \\
    & Security {\footnotesize($\Delta t=3$)} & 6.71 & 13.55 \\
    & Android {\footnotesize($\Delta t=2$)} & 6.77 & 10.68 \\
    & Web {\footnotesize($\Delta t=1$)} & 1.80 & 5.09 \\
    \midrule
    \multirow{4}{*}{$\mathcal{M}_{enc}$} & General {\footnotesize($\Delta t=4$)} & 10.99 & 11.80 \\
    & Security {\footnotesize($\Delta t=3$)} & 23.38 & 22.74 \\
    & Android {\footnotesize($\Delta t=2$)} & 11.15 & 11.91 \\
    & Web {\footnotesize($\Delta t=1$)} & 10.99 & 7.23 \\
    \bottomrule
    \end{tabular*}
\end{table}

\subsubsection*{\textbf{Performance of $\mathcal{M}_{dec}$ and $\mathcal{M}_{enc}$}} In Fig.~\ref{fig:naive_finetuning}, each heatmap depicts the evolution of a metric on the test sets for a single model on one task. The diagonal values in the heatmaps indicate the metric computed on the test set of the current OOD dataset. We observe substantial catastrophic forgetting for both tasks and models and all domains and metrics. That is, we observe a decline of the metrics in all columns, indicating that the model forgets the previous domains when fine-tuned on a new domain. For example, the EM@1 on $\mathcal{D}_{OOD}^1$ (General) drops from $57.37\%$ to $51.73\%$ for $\mathcal{M}_{dec}$. Another example, is the EM on $\mathcal{D}_{OOD}^2$ (Security) dropping from $40.79\%$ to $18.05\%$ for the model $\mathcal{M}_{enc}$. A glance at the heatmaps suggests that the forgetting is more severe for the encoder $\mathcal{M}_{enc}$. Overall, as we increase the number of fine-tuning steps, the forgetting further intensifies in most cases. In addition, for the decoder, the decline in the metrics after one fine-tuning step is less significant compared to the encoder. For example, after one fine-tuning step, the EM@1 on $\mathcal{D}_{OOD}^2$ drops from $60.93\%$ to $57.66\%$ ($-3.27\%$) for the decoder. Whereas it drops from $58.37\%$ to $32.94\%$ ($-25.43\%$) for the encoder. This means that more fine-tuning steps are required for the decoder to forget about past data more severely, whereas, for the encoder, one fine-tuning step is already enough to show a significant decline in performance. This observation confirms our intuition expressed in Section~\ref{sec:models_and_tasks} that $\mathcal{M}_{enc}$ may be less stable than $\mathcal{M}_{dec}$ due to the additional language modeling head randomly initialized.

\subsubsection*{\textbf{Forgetting Metrics}} In Table~\ref{tab:naive-baseline-forgetting}, we calculate the Forgetting metric for the $EM@1$ and $EM$ metrics and for both models. Note that we calculate the $F$ metric at the final time step of the continual fine-tuning. According to the heatmaps of Fig.~\ref{fig:naive_finetuning}, the $F^5$ metric of a domain is the difference between the first and last value of its corresponding column. This difference represents the amount of forgetting that has occurred on each OOD domain during fine-tuning. The $\Delta t$ in the table indicates how recently the model was fine-tuned on a particular domain dataset. We notice that for the decoder $\mathcal{M}_{dec}$, the forgetting is less severe for the EM@1 (used in the API call prediction) than for the EM (used in the API usage prediction). The difference can be attributed to the fact that the API call prediction task is substantially easier than the API usage prediction task. In general, we observe more severe forgetting for the encoder, which further confirms our intuition about the lack of stability of $\mathcal{M}_{enc}$.

Our results and observations illustrate that the problem of forgetting about past data is a major issue for both studied models and significantly more severe for the model $\mathcal{M}_{enc}$. Even with a low number of fine-tuning steps, catastrophic forgetting is already prominent. By considering more fine-tuning steps, we can expect the problem to exacerbate. 

\begin{tcolorbox}[tile,size=fbox,boxsep=2mm,boxrule=0pt,top=0pt,bottom=0pt,
borderline west={1mm}{0pt}{blue!50!white},colback=blue!5!white]
 We conclude that classical transfer learning, the most commonly used fine-tuning method in prior work, is not sufficient and robust enough to allow the model to adapt to new data while retaining knowledge of past data. 
\end{tcolorbox}

\subsection{How Do Continual Learning Approaches Compare to Classical Transfer Learning?}
\label{sec:ft-cl}

\begin{figure}[!t]
    \centering
    \includegraphics[width=\linewidth]{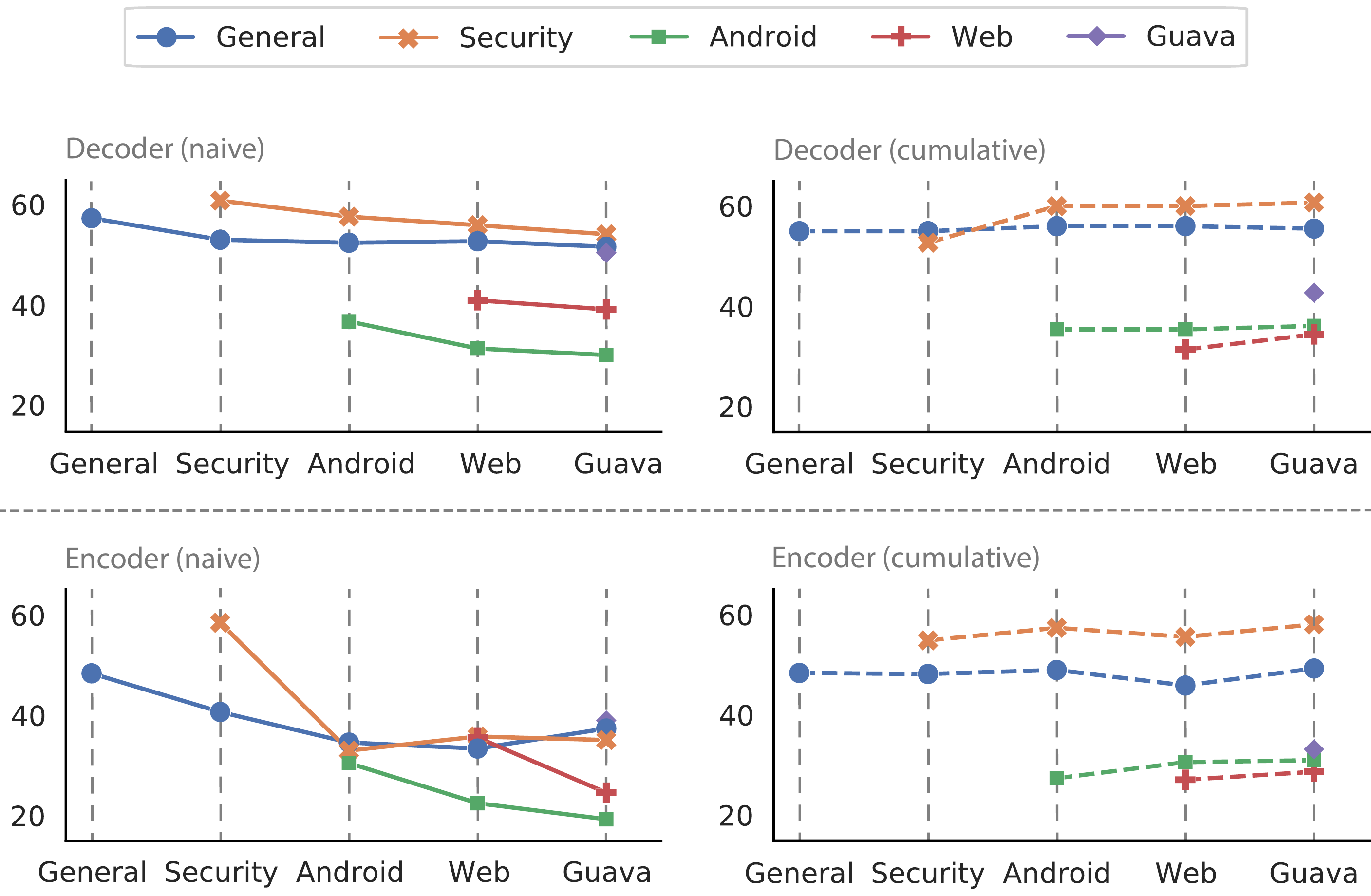}
    \vspace{-2em}
    \caption{Comparison of naive and cumulative fine-tuning settings for both models on API call prediction (EM@1).}
    \label{fig:naive_vs_cumul_call}
\end{figure}

\begin{figure}[!t]
    \centering
    \includegraphics[width=\linewidth]{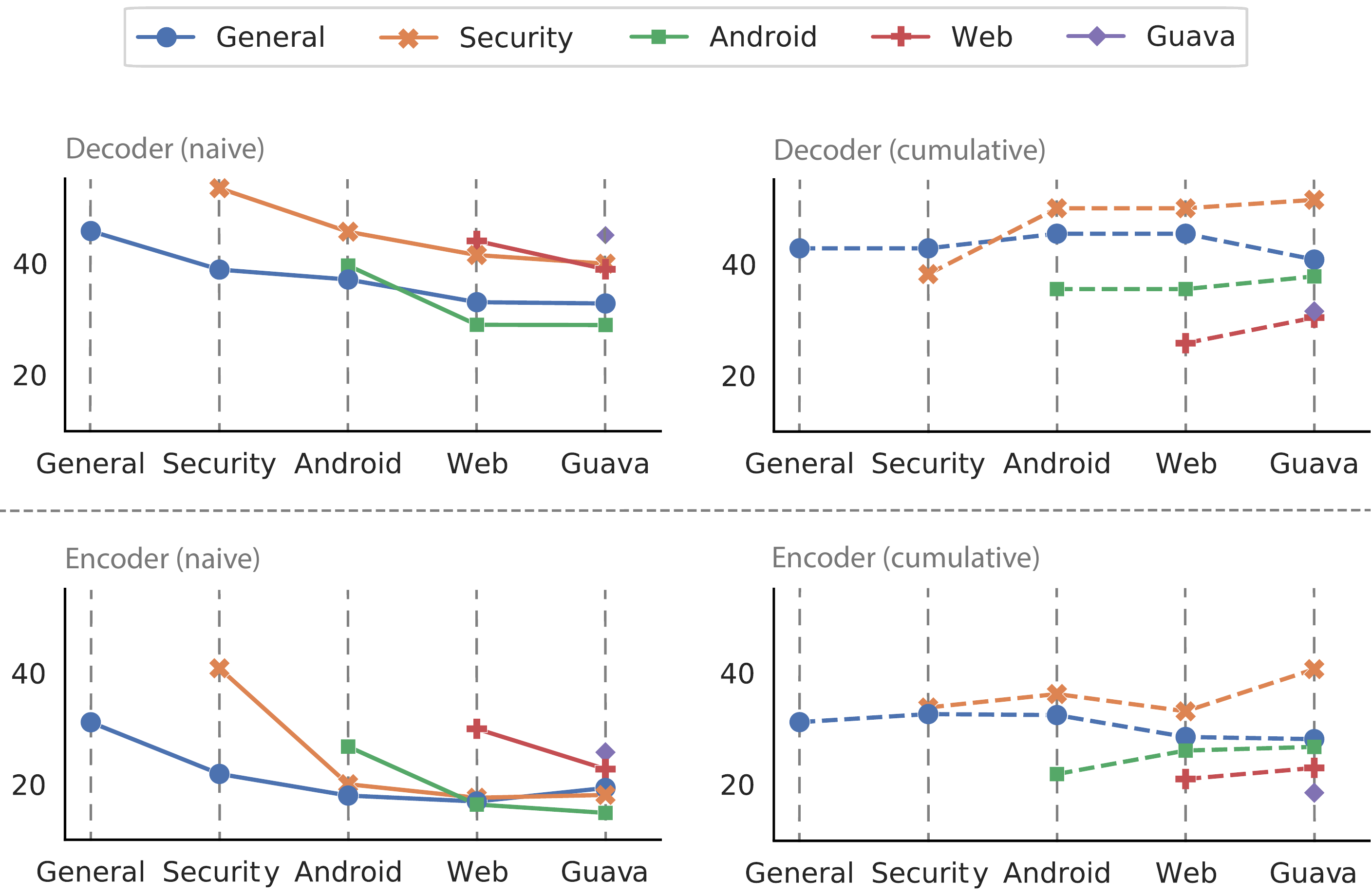}
    \vspace{-2em}
    \caption{Comparison of naive and cumulative fine-tuning settings for both models on API usage prediction (EM).}
    \label{fig:naive_vs_cumul_usage}
\end{figure}

To tackle the problem of catastrophic forgetting highlighted in our previous experiments, we propose to leverage some commonly used continual learning approaches from the literature. In this experiment, the naive fine-tuning approach is the lower-bound baseline, as it has no designed mechanism to mitigate catastrophic forgetting. We begin by introducing an upper-bound approach, referred to as "cumulative fine-tuning", which involves storing all training samples from each OOD training set cumulatively. With this approach, we perform continual fine-tuning using all samples from previous fine-tuning steps in addition to the current ones. This approach is usually upper-bound in continual learning settings as by storing all samples from previous data, the model can optimize its learning to generalize better to the whole stream of data. However, the cumulative fine-tuning approach is not usable in practice for a couple of reasons: (1) we may not always have access to all previous data at any given time, and (2) it requires storing all previous samples and significantly more computations during fine-tuning. This upper-bound approach aims to minimize forgetting while achieving the best overall performance. We compare the cumulative and naive approaches in Fig.~\ref{fig:naive_vs_cumul_call} and Fig.~\ref{fig:naive_vs_cumul_usage}. Next, we introduce additional CL methods, including a replay-based method and three regularization-based methods: EWC~\cite{kirkpatrick2017overcoming}, SI~\cite{zenke2017continual}, and RWalk~\cite{chaudhry2018riemannian}. One advantage of these three methods over the replay method is that they do not require storing samples from previous data while fine-tuning.
We report the Average ($A$) and Forgetting ($F$) metrics for both tasks and models on the EM@1 and EM metrics in Table~\ref{tab:cl-baselines_call} and Table~\ref{tab:cl-baseline_usage}. Note that there is no Forgetting metric for Guava as it is the last domain the PLMs are fine-tuned on.

\subsubsection*{\textbf{Fine-Tuning Details}} We use the same fine-tuning procedure as in the previous experiment. For the replay baseline, we set the buffer size to 200, \textit{i.e.,} number of sampled stored from past OOD training sets. We provide all our hyperparameters and further details about the implementations in our replication package.

\renewcommand{\arraystretch}{1.2}
\setlength{\arrayrulewidth}{.5pt}
\begin{table*}[!t]
\centering
\caption{Continual learning approaches results for API call prediction using the EM@1 metric.} \label{tab:cl-baselines-call}
\vspace{-1em}
\resizebox{1.5\columnwidth}{!}{%
    \begin{tabular*}{\linewidth}{ll@{\extracolsep{\fill}}*{10}{c}}
    \toprule
    & & \multicolumn{2}{c}{General} & \multicolumn{2}{c}{Security} & \multicolumn{2}{c}{Android} & \multicolumn{2}{c}{Web} & \multicolumn{2}{c}{Guava} \\ 
    \cline{3-12} 
    \textsc{Model} & \textsc{Method} & $A_{\text{EM@1}}$ $\uparrow$  & $F^5_{\text{EM@1}}$ $\downarrow$ & $A_{\text{EM@1}}$ & $F^5_{\text{EM@1}}$ & $A_{\text{EM@1}}$ & $F^5_{\text{EM@1}}$ & $A_{\text{EM@1}}$ & $F^5_{\text{EM@1}}$ & $A_{\text{EM@1}}$ & $F^5_{\text{EM@1}}$ \\ 
    \midrule
    \multirow{7}{*}{$\mathcal{M}_{dec}$} & Naive & 53.49 & 5.64 & 57.21 & 6.71 & 32.75 & 6.77 & 40.06 & 1.80 & \colorbox{gray!50}{\textbf{50.47}} & -- \\
    \cdashline{2-12}
    & EWC~\cite{kirkpatrick2017overcoming} & 53.22 & 7.02 & 57.16 & 7.49 & 33.73 & 5.72 & \colorbox{gray!30}{40.14} & 3.77 & 49.59 & -- \\
    & SI~\cite{zenke2017continual} & 54.65 & 3.57 & \colorbox{gray!50}{\textbf{59.24}} & 3.45 & 34.04 & 2.39 & 38.93 & \colorbox{gray!30}{1.36} & 48.16 & -- \\
    & RWalk~\cite{chaudhry2018riemannian} & 54.38 & \colorbox{gray!30}{2.39} & 57.39 & \colorbox{gray!30}{2.80} & 31.64 & \colorbox{gray!30}{1.97} & 38.19 & 1.65 & 45.28 & -- \\
    & Replay & \colorbox{gray!50}{\textbf{55.66}} & 4.41 & \colorbox{gray!30}{58.87} & 2.98 & \colorbox{gray!30}{34.66} & 2.01 & \colorbox{gray!50}{\textbf{41.12}} & 2.41 & \colorbox{gray!30}{49.72} & -- \\
    & Cumulative & \colorbox{gray!30}{55.63} & \colorbox{gray!50}{\textbf{-0.51}} & 58.44 & \colorbox{gray!50}{\textbf{-8.02}} & \colorbox{gray!50}{\textbf{35.74}} & \colorbox{gray!50}{\textbf{-0.73}} & 32.99 & \colorbox{gray!50}{\textbf{-3.01}} & 42.79 & -- \\
    \midrule
    \midrule
    \multirow{7}{*}{$\mathcal{M}_{enc}$} & Naive & 38.78 & 10.99 & 40.49 & 23.38 & 24.01 & 11.15 & \colorbox{gray!30}{30.05} & 10.99 & 38.85 & -- \\
    \cdashline{2-12}
    & EWC~\cite{kirkpatrick2017overcoming} & 39.38 & 9.84 & 44.10 & 22.15 & 23.93 & 10.58 & 29.22 & 7.53 & \colorbox{gray!50}{\textbf{40.66}} & -- \\
    & SI~\cite{zenke2017continual} & 44.29 & 5.94 & 50.05 & \colorbox{gray!30}{8.10} & 21.39 & \colorbox{gray!30}{4.02} & 27.79 & \colorbox{gray!30}{2.56} & 35.67 & -- \\
    & RWalk~\cite{chaudhry2018riemannian} & 43.42 & 6.07 & 48.05 & 14.74 & 22.23 & 7.10 & 29.75 & 4.37 & 36.10 & -- \\
    & Replay  & \colorbox{gray!30}{45.15} & \colorbox{gray!30}{5.48} & \colorbox{gray!30}{51.56} & 10.56 & \colorbox{gray!30}{24.31} & 8.27 & \colorbox{gray!50}{\textbf{32.53}} & 3.92 & \colorbox{gray!30}{40.22} & -- \\
    & Cumulative & \colorbox{gray!50}{\textbf{48.06}} & \colorbox{gray!50}{\textbf{-0.92}} & \colorbox{gray!50}{\textbf{56.40}} & \colorbox{gray!50}{\textbf{-3.15}} & \colorbox{gray!50}{\textbf{29.59}} & \colorbox{gray!50}{\textbf{-3.62}} & 27.79 & \colorbox{gray!50}{\textbf{-1.65}} & 33.10 & -- \\
    \bottomrule
    \end{tabular*}
    }
    \label{tab:cl-baselines_call}
\end{table*}

\renewcommand{\arraystretch}{1.2}
\setlength{\arrayrulewidth}{.5pt}
\begin{table*}[!t]
\centering
\caption{Continual learning approaches results for API usage prediction using the EM metric.} \label{tab:cl-baselines-usage}
\vspace{-1em}
\resizebox{1.5\columnwidth}{!}{%
    \begin{tabular*}{.9\linewidth}{ll@{\extracolsep{\fill}}*{10}{c}}
    \toprule
    & & \multicolumn{2}{c}{General} & \multicolumn{2}{c}{Security} & \multicolumn{2}{c}{Android} & \multicolumn{2}{c}{Web} & \multicolumn{2}{c}{Guava} \\ 
    \cline{3-12} 
    \textsc{Model} & \textsc{Method} & $A_{EM}$ $\uparrow$  & $F_{EM}^5$ $\downarrow$ & $A_{EM}$ & $F_{EM}^5$ & $A_{EM}$ & $F_{EM}^5$ & $A_{EM}$ & $F_{EM}^5$ & $A_{EM}$ & $F_{EM}^5$ \\ 
    \midrule
    \multirow{7}{*}{$\mathcal{M}_{dec}$} & Naive & 37.32 & 13.00 & 44.96 & 13.55 & 32.31 & 10.68 & \colorbox{gray!50}{\textbf{41.30}} & 5.09 & \colorbox{gray!30}{44.87} & -- \\
    \cdashline{2-12}
    & EWC~\cite{kirkpatrick2017overcoming} & 36.88 & 12.95 & 44.84 & 13.08 & \colorbox{gray!30}{33.92} & 9.46 & 39.00 & 6.73 & \colorbox{gray!50}{\textbf{45.71}} & -- \\
    & SI~\cite{zenke2017continual} & 40.36 & 8.26 & \colorbox{gray!50}{\textbf{49.58}} & 6.89 & 30.01 & 3.24 & 36.95 & \colorbox{gray!30}{1.65} & 43.14 & --  \\
    & RWalk~\cite{chaudhry2018riemannian} & \colorbox{gray!30}{40.43} & \colorbox{gray!30}{6.23} & 47.11 & \colorbox{gray!30}{4.04} & 33.34 & \colorbox{gray!30}{2.63} & 36.54 & 2.13 & 41.22 & -- \\
    & Replay & 39.49 & 11.11 & 46.88 & 8.21 & 33.39 & 7.63 & \colorbox{gray!30}{39.49} & 6.08 & 43.65 & -- \\ 
    & Cumulative & \colorbox{gray!50}{\textbf{43.29}} & \colorbox{gray!50}{\textbf{2.02}} & \colorbox{gray!30}{47.26} & \colorbox{gray!50}{\textbf{-13.33}} & \colorbox{gray!50}{\textbf{36.09}} & \colorbox{gray!50}{\textbf{-2.28}} & 27.92 & \colorbox{gray!50}{\textbf{-4.59}} & 31.35 & --  \\
    \midrule
    \midrule
    \multirow{7}{*}{$\mathcal{M}_{enc}$} & Naive & 21.41 & 11.80 & 24.09 & 22.74 & 19.30 & 11.91 & \colorbox{gray!50}{\textbf{26.32}} & 7.23 & 25.71 & -- \\
    \cdashline{2-12}
    & EWC~\cite{kirkpatrick2017overcoming} & 21.32 & 11.53 & 26.36 & 21.02 & \colorbox{gray!30}{19.43} & 11.96 & \colorbox{gray!30}{25.74} & 8.38 & \colorbox{gray!50}{\textbf{28.74}} & -- \\
    & SI~\cite{zenke2017continual} & \colorbox{gray!30}{27.22} & \colorbox{gray!30}{5.03} & \colorbox{gray!30}{30.85} & \colorbox{gray!30}{8.23} & 18.57 & \colorbox{gray!30}{2.20} & 23.03 & \colorbox{gray!30}{1.65} & 21.26 & -- \\
    & RWalk~\cite{chaudhry2018riemannian} & 25.21 & 8.80 & 29.25 & 12.23 & 19.10 & 7.62 & 25.00 & 4.28 & 24.23 & -- \\
    & Replay & 23.48 & 13.54 & 29.94 & 13.96 & 18.09 & 11.88 & 24.51 & 5.92 & \colorbox{gray!30}{26.48} & -- \\
    & Cumulative & \colorbox{gray!50}{\textbf{30.50}} & \colorbox{gray!50}{\textbf{3.05}} & \colorbox{gray!50}{\textbf{35.89}} & \colorbox{gray!50}{\textbf{-6.88}} & \colorbox{gray!50}{\textbf{24.81}} & \colorbox{gray!50}{\textbf{-4.88}} & 21.88 & \colorbox{gray!50}{\textbf{-1.97}} & 18.43 & --  \\
    \bottomrule
    \end{tabular*}
    }
    \label{tab:cl-baseline_usage}
\end{table*}

\subsubsection*{\textbf{Cumulative Fine-Tuning}} In Fig.~\ref{fig:naive_vs_cumul_call}, we compare the naive and cumulative approaches for the API call prediction task ($EM@1$) on both decoder and encoder models. Each curve illustrates the evolution of the EM@1 on a particular OOD test set. The figure further demonstrates how the naive approach (bottom-left part of the figure) with the encoder leads to significantly more forgetting than for the decoder, as previously discussed. At the left of Fig.~\ref{fig:naive_vs_cumul_call}, we observe that the cumulative fine-tuning approach effectively eliminates the catastrophic forgetting issue for both models. Specifically, the EM@1 does not decrease over time and even increases throughout the fine-tuning, indicating improvement during continual fine-tuning, also known as positive transfer. In Fig.~\ref{fig:naive_vs_cumul_usage}, we make the same observations for the API usage prediction task on the EM metric.

\subsubsection*{\textbf{Continual Learning Approaches}} Table~\ref{tab:cl-baselines_call} reports the Average and Forgetting metrics of the EM@1 on each OOD test set for $\mathcal{M}_{dec}$ and $\mathcal{M}_{enc}$, with the naive fine-tuning approach as baseline. Similarly to Section~\ref{sec:ft-naive}, we compute the $F$ metric at the end of the continual fine-tuning. Firstly, we observe that for both models, the cumulative fine-tuning approach is the best option to mitigate catastrophic forgetting and generally leads to the best $A_{EM@1}$. With the cumulative approach, the $F_{EM@1}^5$ metric is always negative, which indicates a positive transfer (an increase in the EM@1). For instance, we get $-8.02$ in $F_{EM@1}^5$ for $\mathcal{M}_{dec}$ in the Security domain, \textit{i.e.,} an increase of $+8.02$ in the metric through fine-tuning. However, we observe large gaps between the $A_{EM@1}$ obtained using the cumulative approach and the naive approach on the Guava dataset (last fine-tuning step). We hypothesize that with an ever-increasing replay buffer, the models can no longer learn from new data and thus lose their ability to adapt with time. In addition to being computationally intensive, the cumulative fine-tuning approach is not scalable and robust, as previously mentioned. Overall, all other CL approaches, except EWC, greatly reduce forgetting and show a superior average EM@1 compared to the naive approach. The Replay approach generally produces the best or second best $A_{EM@1}$. Without the cumulative approach, RWalk is the best method to mitigate forgetting for $\mathcal{M}_{dec}$, whereas SI is better for $\mathcal{M}_{enc}$. In Table~\ref{tab:cl-baseline_usage}, we report the results for the API usage prediction task. We observe similar trends, except that the Replay approach is less effective for both models. However, RWalk and SI are the best methods for $\mathcal{M}_{dec}$ and $\mathcal{M}_{enc}$, respectively.

\begin{tcolorbox}[tile,size=fbox,boxsep=2mm,boxrule=0pt,top=0pt,bottom=0pt,
borderline west={1mm}{0pt}{blue!50!white},colback=blue!5!white]
In this final experiment, we demonstrate that continual learning methods, including two replay-based methods (Replay and Cumulative) and two regularization-based methods (SI and RWalk) effectively reduces catastrophic forgetting while achieving similar or superior effectiveness compared to classical transfer learning on both tasks.
\end{tcolorbox}


\section{Discussion}
\label{sec:discussion}
In this section, we address some threats to the validity of our study. We then discuss the broader impact of our study and various opportunities for future work.

\subsection{Threats to Validity}
\subsubsection*{\textbf{Threats to External Validity}} 
We identified a main threat regarding the monolingual aspect of our dataset. 
Our OOD scenario requires extracting API usage sequences from the source code. 
Therefore, integrating more programming languages demands substantial additional effort, which we deliberately leave for future work. 
In addition, the construction of our dataset does not include any programming language-specific design and avoids any data leakage between the ID and OOD data. 
Consequently, it is highly likely that our results are not affected by the programming language of the data.

Another threat related to the data is the choice of the OOD domains and APIs. 
To mitigate this threat, we selected five domains covering different types of programs. 
Specifically, we selected 10 random interfaces per domain. Our results show that catastrophic forgetting is observed consistently for all domains, and the selection of different interfaces would result in different intensities in forgetting. We leave the study of this qualitative aspect for future work. 

The choice of the downstream tasks presents another external threat to validity of our study.
We employed two generation tasks, API call and API usage prediction.
We focus on APIs-related tasks because APIs are an important part of the distribution of code tokens in programs and give lots of information about the semantics of programs.
We observe significant catastrophic forgetting in these two API-related tasks and hypothesize that catastrophic forgetting could appear in other SE tasks because of the importance of APIs in code. For instance, previous work found that APIs play important roles in writing the summarization of code~\cite{hu2018summarizing}, detecting code clones~\cite{nafi2019clcdsa}, retrieving code given a query~\cite{lv2015codehow}, etc.
We leave the investigation of the OOD phenomenon in other tasks as future work.

We identified an external threat to validity related to the limited number of fine-tuning steps in our continual fine-tuning settings. In practice, a PLM deployed to a real production environment would potentially face a larger number of fine-tuning steps throughout its lifetime. In this paper, we showed that both PLMs suffer from severe catastrophic forgetting, although we only consider five fine-tuning steps. We also demonstrated that more steps generally result in more forgetting about past data.

Finally, the selection of the size of the PLMs, in terms of the number of trainable parameters, constitutes a potential threat to the validity of our study. 
While increasing the number of parameters may still result in OOD generalization issues due to the design of our datasets, it is uncertain whether catastrophic forgetting would occur with the same magnitude for larger models. 
Our experiments were performed under limited computational resources, which required us to consider architectures with a limited number of parameters.
To mitigate this threat, we maximized the size of the models considering our limited resources. We pre-train PLMs with 110M and 125M parameters which are within the range of PLMs such as CodeBERT~\cite{feng2020codebert}, CodeT5~\cite{wang2021codet5} or CodeGPT~\cite{lu2021codexglue}.

\subsubsection*{\textbf{Threats to Internal Validity}} The hyperparameter choices for our CL approaches constitute the main threat to internal validity. We selected our hyperparameters based on values used in prior works about continual learning~\cite{jie2022alleviating, kirkpatrick2017overcoming, chaudhry2018riemannian, zenke2017continual}. These hyperparameters can be optimized for our scenario by using search methods, which tend to have a high computational cost. However, this aspect is not critical to the study as we have already shown the advantages of incorporating continual learning techniques with reasonable hyperparameter values.

\subsubsection*{\textbf{Threats to Construct Validity}} We identified one threat to construct validity related to the choice of our evaluation metrics. We mitigate this threat by selecting metrics widely used in prior works to evaluate code generation tasks~\cite{xu2022systematic, ren2020codebleu}. Additionally, we adapted continual learning metrics from prior works~\cite{chaudhry2018riemannian, jie2022alleviating} to evaluate our continual fine-tuning scenario.

\subsection{Broader Impact and Opportunities}
\label{sec:broader_impact}
Our study sheds light on the performance of PLMs of code in a continual learning setting for out-of-distribution generalization. 
We believe that this initial exploration of continual learning for code ({\em CL4Code}) will inspire further investigation in this important area. 
Our findings highlight two potential areas for future research: improving dataset and benchmark creation, and expanding the application of CL4Code to a wider range of use cases.

\subsubsection*{\textbf{Datasets and Benchmarks}}
Our findings in Section~\ref{sec:zero-shot} highlight a substantial disparity in the performance of a PLM between ID and OOD data. 
Our results, along with a previous work~\cite{zhou2021assessing}, indicate that evaluating PLMs on ID data often leads to inflated metrics and results in overly optimistic conclusions in terms of the performance.
Therefore, it is crucial to develop OOD datasets for code in order to evaluate the real-world generalizability of PLMs, as previously emphasized~\cite{zhou2021assessing, yang2022glue}. 
Moreover, aligning dataset designs with continual learning scenarios offers the potential to evaluate the PLM's ability to adapt to changing environments, which is crucial for practical deployment.

Improving benchmarks for PLMs of code is another promising direction for future research. 
Benchmarks such as CodeXGlue~\cite{lu2021codexglue}  play a crucial role by providing standardized evaluations of models of code and enabling reproducible experimental results.
However, as such researches progress at a rapid pace, widely used benchmarks often become outdated quickly. 
In particular, Kiela et al.~\cite{kiela2021dynabench} showed that benchmarks such as GLUE~\cite{wang2018glue} in NLP saturate, meaning the milestones set by the benchmark are reached.
Thus, continued efforts to enhance benchmarks in deep learning for code are vital in establishing concrete goals and driving research to enhance the performance of the models being evaluated.
Recently, Yang et al.~\cite{yang2022glue} proposed GLUE-X, a comprehensive benchmark consisting of 13 datasets to test PLMs on OOD data across eight NLP tasks. 
The benchmark includes OOD datasets that are distinct from those in the original GLUE benchmark.
Developing OOD benchmarks for code similar to GLUE-X~\cite{yang2022glue} would greatly contribute to the growth of research on OOD generalization for PLMs of code. 
One potential approach is to compile a new set of OOD datasets that are not included in the existing CodeXGlue benchmark, and use them to test PLMs of code.
Furthermore, exploring the design of OOD scenarios specific to software changes, as demonstrated in the present study, can provide a valuable foundation for future code benchmark initiatives. Our dataset and methodology for extracting OOD samples for API evolution scenarios can serve as a starting point for these endeavors.

\subsubsection*{\textbf{Continual Learning for Code}}
Our findings in Section~\ref{sec:ft-naive} highlight the challenge of catastrophic forgetting that PLMs of code encounter in a continual fine-tuning scenario with OOD data.
Our study serves as a starting point for exploring the adaptability of PLMs of code in a variety of continual learning scenarios.
For instance, these scenarios can be based on domain adaptation, where PLMs must adapt to new kinds of data such as new, unseen programming languages or code repositories as discussed in prior studies~\cite{koh2021wilds, hajipour2022simscood, hu2022codes}. 
Additionally, incorporating continual learning into a multi-task learning framework is highly relevant to software engineering, given the multitude of downstream tasks involved. 

In Section~\ref{sec:ft-cl}, our results demonstrate the effectiveness of continual learning methods in mitigating catastrophic forgetting in PLMs of code.
We chose to explore these widely used methods as a first step in the research on continual learning for code.
In the future, more sophisticated techniques from NLP, as discussed in Section~\ref{sec:rl_cl}, can be evaluated.
Furthermore, the creation of continual learning methods specifically tailored to source code has the potential to further reduce catastrophic forgetting in PLMs of code.

Finally, we did not focus our study on large language models (LLMs) as it would require a tremendous amount of available computational resources to pre-train an LLM from scratch under our OOD scenario. 
Nonetheless, we foresee that continuously adapting LLMs to new emerging datasets and benchmarks constitutes an exciting avenue for future work. 
In this context, and as fully fine-tuning LLMs is computationally costly, we believe that combining continual learning with parameter-efficient fine-tuning (PEFT) techniques might be beneficial to further enhance the capabilities of LLMs. 
These 
PEFT techniques have already shown promising results in LLMs for code intelligence~\cite{wang2022no, choi2023codeprompt, weyssow2023exploring}.

\section{Related Work}
\label{sec:related_work}
\subsection{Out-Of-Distribution Generalization}

\subsubsection*{\textbf{Natural Language Processing}} Recent studies have revealed that PLMs are susceptible to generating inaccurate predictions when encountering OOD data~\cite{hendrycks-etal-2020-pretrained, shi2022how}. In NLP, this issue can manifest itself in situations where the domain of the test data differs from the pre-training data~\cite{gururangan-etal-2020-dont}. One approach to addressing this problem is to fine-tune PLMs on domain-specific datasets using efficient transfer learning techniques. For example, \cite{ruder-etal-2019-transfer, howard-ruder-2018-universal} demonstrated that such approaches help PLMs in learning domain-specific knowledge and improve their generalization to unseen domains. Additionally, new datasets and benchmarks allow for further research on PLM domain adaptation. For instance, Williams et al.~\cite{williams2017broad} introduced the MultiNLI dataset, containing text data from a variety of domains for PLM domain adaptation. Conneau et al.~\cite{conneau2018xnli} proposed a cross-lingual NLI dataset for evaluating the cross-lingual transferability of PLMs. Recently, Yang et al.~\cite{yang2022glue} introduced GLUE-X, a benchmark for evaluating PLMs' ability to generalize to OOD data.

\subsubsection*{\textbf{Deep Learning for Code}} The study of OOD generalization of PLMs of code is an emerging research area. Assessing their generalizability and designing efficient techniques to improve their robustness to OOD scenarios is essential for the practical usability of PLMs of code~\cite{zhou2021assessing}. Previous work in this field has focused on designing OOD datasets that simulate specific distribution shifts of program data. Koh et al.~\cite{koh2021wilds} presented PY150-Wilds, a Python dataset in which the test data consists of code repositories not appearing in the training data. The authors demonstrated performance gaps between the model on ID and OOD data. However, it is important to note that while the design choice is sound, it may not reflect strong OOD phenomena as the distribution of code tokens across different repositories may still be highly similar. More recently, Hu et al.~\cite{hu2022codes} proposed a benchmark to evaluate the performance of code models under different distribution shift scenarios, including programmer, time, or token distribution shifts. In their study, the authors found that PLMs such as CodeBERT were robust against distribution shifts. However, they demonstrated that on a simple classification task with small datasets. In addition, the authors did not control the pre-training data of the studied PLMs, which can result in important data leakage between the pre-training and OOD test data. This problem of data leakage is critical as some of the test data may have been seen by the model during pre-training. Overall, this is a prime threat to the validity of the OOD scenario that may lead to obtaining inflated metrics on the OOD test data. Finally, Hajipour et al.~\cite{hajipour2022simscood} analyzed the performance of PLMs of code on a syntax-based, semantic-based and complexity-based OOD scenario and highlighted that the models exhibit poor generalizability when faced with OOD samples. However, it is important to point out that the OOD scenarios used in this study may be too artificial. For instance, in the syntax-based scenario, some language-specific tokens are masked at training to study how the model generalizes to unseen language tokens. Such a scenario is unrealistic as it does not reflect the nature of OOD data that a PLM of code is likely to encounter in the real world. Additionally, there is no practical motivation for masking specific tokens while training the model.

In this study, we propose an OOD dataset that accurately represents the dynamic nature of software codebases in the real world. Specifically, we focus on the scenario where a PLM must adapt to new, unseen APIs over time, a well-established problem in the literature~\cite{nita2010using, proksch2016evaluating}. To ensure the validity of our experiments, we thoroughly control our PLM setup to prevent any data leakage between the pre-training, fine-tuning, and test data. This allows us to create an OOD generalization scenario that is as close to reality as possible, an aspect that has been overlooked in previous works.

\subsection{Continual Learning for Pre-trained Language Models}
\label{sec:rl_cl}

Continual learning has been studied to adapt pre-trained language models based on the Transformer architecture~\cite{vaswani2017attention} to new domains or tasks in NLP. For example, Cao et al.~\cite{cao2020incremental} proposed a method to continually learn from new classes of events in textual data to detect them without degradation of the accuracy over time. Douillard et al.~\cite{douillard2022dytox} introduced DyTox, a method that utilizes an encoder-decoder transformer for multiple tasks by expanding the network with task-specific special tokens, allowing for continual learning of new tasks with a low computational and memory footprint. Ermis et al.~\cite{ermis2022memory} proposed a memory-efficient approach for transformers to continually learn new tasks by sharing information across tasks and expanding the network with task-specific modules. Similarly, Vladymyrov et al.~\cite{vladymyrov2023continual} proposed the HyperTransformer architecture to continually learn new tasks by generating task-specific convolutional neural network weights in a few-shot learning setting and updating the task-specific weights to avoid catastrophic forgetting. Lastly, Jie et al.~\cite{jie2022alleviating} leverage continual learning to avoid representational shifts in PLMs by proposing a new hierarchical fine-tuning method that prevents excessive changes in the representation spaces of the neural network in a continual fine-tuning setting. 

Recent advances in NLP highlight the crucial need for PLMs to adapt to changing environments and maintain their performance on new data and tasks. In the field of software engineering, the application of continual learning to PLMs of code is essential for developing methods that enable the model to robustly adapt to new codebases and tasks over time. 
To the best of our knowledge, only a couple of prior studies utilized continual learning in the context of code intelligence.
Baudry et al.~\cite{baudry2021software} demonstrate the benefits of leveraging continual learning to fix bugs when considering a continuous stream of code change with continuous integration development platforms. 
The scope of our study differ from this prior work in many aspects. 
First, contrary to this prior work, our study focuses on PLM architectures which broadens the potential applicability of our approach to a broader range of tasks. 
Secondly, we compare numerous continual learning techniques in our OOD scenario with PLMs, whereas this previous work only consider a continual learning scenario without leveraging continual learning techniques such as replay buffer or EWC.
More recently, Gao et al.~\cite{gao2023keeping} made similar findings than ours by showing that PLMs suffer from catastrophic forgetting in continual learning scenarios and that replay-based approaches allow to effectively mitigate forgetting.
We believe that these prior works and our study break new ground by introducing the first approaches on the utilization of continual learning for PLMs of code.

\section{Conclusion and future work}
\label{sec:ccl}
Our study exposes the limitations of pre-trained language models of code in handling out-of-distribution data in a continual fine-tuning scenario. 
Our results reveal that OOD data significantly decreases the PLMs' effectiveness in two API-related downstream tasks compared to ID data.
Our findings indicate that classical transfer learning fails to adapt the PLMs to new, unseen APIs in this evolution scenario.
Additionally, we observe instances of catastrophic forgetting, prompting us to explore methods that address this issue. 
In our final experiments, we demonstrate that replay-based and regularization-based continual learning techniques can effectively mitigate catastrophic forgetting while retaining or enhancing the performance of the PLMs in both downstream tasks.
In future work, we intend to explore more OOD scenarios to further evaluate the generalizability of PLMs of code and develop relevant OOD generalization benchmarks for code.
Additionally, we plan to implement more advanced continual learning methods tailored to source code to enhance the adaptability of PLMs of code. Finally, we aim to investigate OOD detection methods to automatically identify OOD data in PLMs, thereby improving their performance.

\section*{Data Availability}
\label{sec:data_availability}
We publicly release all the code, data and models to reproduce the experiments of our study.
The following repository contains instructions on how to acquire the data and pre-train, fine-tune and test the PLMs:
\url{https://github.com/martin-wey/cl-code-apis}


\balance
\bibliographystyle{ACM-Reference-Format}
\bibliography{references}


\end{document}